\documentclass[conference]{IEEEtran}
\usepackage{amsmath,amssymb,amsfonts}
\usepackage{algorithm}
\usepackage{algorithmic}
\usepackage{graphicx}
\usepackage{textcomp}
\usepackage{multirow}
\usepackage{pstricks, pstricks-add}
\usepackage{pgf}
\usepackage{pgfplots}
\usepackage{amsmath}
\usepackage{subcaption}
\usepackage[
skip=0.3\baselineskip
]{caption}
\usepackage[hidelinks]{hyperref}
\usepackage[usestackEOL]{stackengine}
\usepackage{booktabs}       

\def\BibTeX{{\rm B\kern-.05em{\sc i\kern-.025em b}\kern-.08em
		T\kern-.1667em\lower.7ex\hbox{E}\kern-.125emX}}

\usepackage{pifont}

\everymath=\expandafter{\the\everymath\displaystyle}
\IfFileExists{scrextend.sty}{
	\usepackage[fontsize=10.000000pt]{scrextend}
}{
	\renewcommand{\normalsize}{\fontsize{10.000000}{12.000000}\selectfont}
	\normalsize
}

\makeatletter\@ifpackageloaded{underscore}{}{\usepackage[strings]{underscore}}\makeatother

\begin{document}
	
	\title{Carbon-Aware Temporal Data Transfer Scheduling Across Cloud Datacenters
	}
	
	\author{\IEEEauthorblockN{Elvis Rodrigues, Jacob Goldverg, Tevfik Kosar}
		\IEEEauthorblockA{Department of Computer Science and Engineering \\
			University at Buffalo (SUNY), Amherst, NY 14260, USA\\
			Email: \{elvisdav, jacobgol, tkosar\}@buffalo.edu}
	}
	
	\pagestyle{plain}
	\maketitle
	
	\begin{abstract}
		Inter-datacenter communication is a significant part of cloud operations and produces a substantial amount of carbon emissions
		for cloud data centers, where the environmental impact has already been a pressing issue. In this paper, we present a novel carbon-aware temporal data transfer scheduling framework, called LinTS, which promises to significantly reduce the carbon emission of data transfers between cloud data centers. LinTS produces a competitive transfer schedule and makes scaling decisions, outperforming common heuristic algorithms. LinTS can lower carbon emissions during inter-datacenter transfers by up to 66\% compared to the worst case and up to 15\% compared to other solutions while preserving all deadline constraints.
	\end{abstract}
	
	
	\section{Introduction}
	Demand for power in the U.S. is projected to grow by 2.4\% over the next 7 years, of which datacenters are a significant component with a 15\% projected growth till 2030 \cite{GoldmanSachs}. Much of this demand today stems from AI training and inference with large AI models, emerging new cloud services and network activity, and diminishing gains of power efficiency with new iterations of datacenter hardware. In response, technology companies and cloud providers have set ambitious targets to achieve net-zero, and even net-negative, carbon emissions by 2030 \cite{GoogleSust24, MetaSust24, MicroSust24}. With increasing demand from AI workloads, significant investments have also been made in emerging green power generation methods such as Small Modular Reactors (SMRs) \cite{GoogleKairos}. However, despite these investments, a large portion of the energy consumed by cloud datacenters is generated from brown, carbon-intense sources due to the lack of reliable renewable energy supply, undermining decarbonization efforts worldwide \cite{GoogleSust24, CarbonExplorer}. Alongside efforts to make datacenter hardware more power efficient, there have been many studies into the nature and trends of datacenter workloads and strategies to make these workloads more carbon efficient \cite{onTheLimitations, MetricsForSus, mosaic, Treehouse, googleCarbon}.

Among all the workloads executed by a cloud provider, a significant amount of time and resources are spent transferring datasets between different regions and managing these datasets to maximize availability for clients and users while minimizing costs to providers. Global inter-datacenter traffic has exceeded 1.4 Zettabytes and is projected to grow by 30\% every year \cite{CiscoReport}. 
This surge in global inter-datacenter traffic has also made the energy consumption and carbon footprint of data transfers a critical concern for cloud datacenters, where the environmental impact has already been a pressing issue.
The datacenters are projected to consume over 500 TWh of energy in 2025, emitting roughly 225 metric megatons of CO$_2$ calculated from the global average carbon intensity of 450 gCO$_2$ per kWh \cite{emberenergyGlobalElectricity}, and serious efforts are needed to curb the emissions of inter-datacenter communication. 
Although networking technologies have advanced significantly in recent years, data transfers over networks remain highly energy-intensive and contribute substantially to carbon emissions. A study reported that sending hard drives between remote institutions via airplanes is much less carbon-emitting than transferring the data over communication networks~\cite{aujoux2021estimating}. 

Given the urgency of reducing carbon emissions of cloud datacenters, cloud providers must now also consider minimizing the carbon footprint of the data transfer tasks. There are several strategies at the disposal of cloud providers and application developers to achieve this, starting with the sustainable design of the datacenter itself with factors such as the embodied carbon cost and failure rate of hardware, energy suppliers, and cooling solutions to consider \cite{MetricsForSus}. Cyclical variations and regional differences in the sources of the power grid's energy supply can be studied to exploit temporal and spatial opportunities to schedule data transfers and place dataset replicas to further reduce carbon emissions. For this purpose, tools like $Electricity Maps$ \cite{ElectricityMaps} and $WattTime$ \cite{WattTime} have become widely popular. Inter-datacenter transfers are often time-flexible and interruptible and thus can benefit from careful scheduling and preemption \cite{CarbonScaler}. For instance, a study shows that 91\% of all inter-datacenter traffic at Baidu's datacenters is replication-related and delay-tolerant, which is considered to corroborate the traffic pattern of other large-scale cloud service providers \cite{BDS}. 

\begin{figure*}[t]
	\centering
	\begin{subfigure}[b]{0.99\columnwidth}  
		\includegraphics[width=\linewidth]{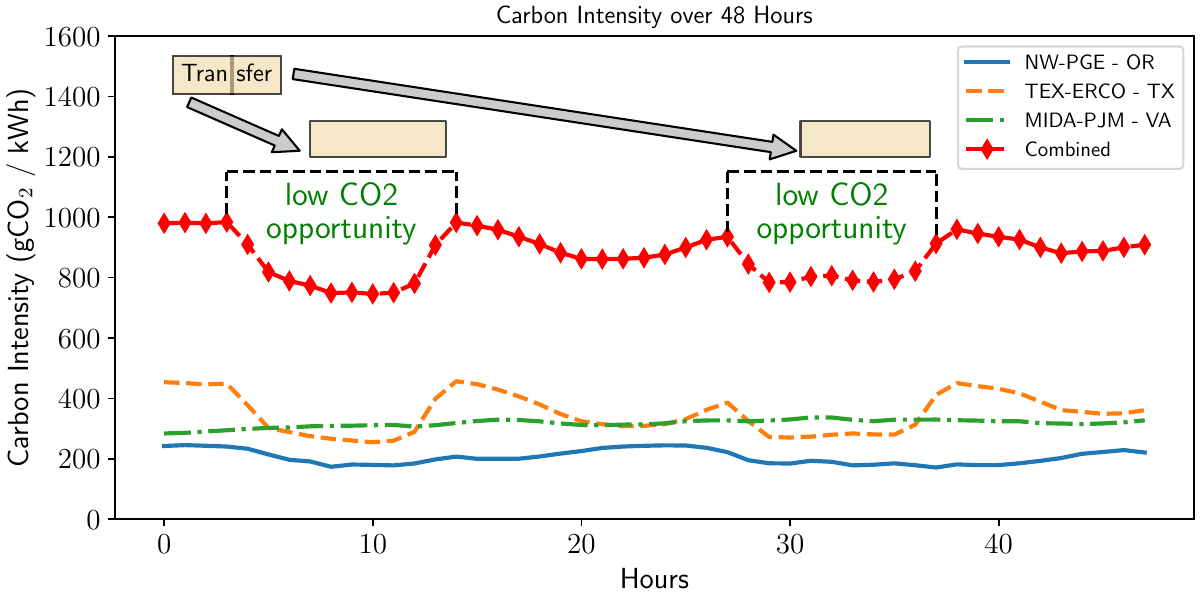}
		\caption{Portioning and scheduling of data transfers over 48 hours with source, intermediate, and destination sites.}
	\end{subfigure}
	\hspace{0.04\columnwidth}  
	\begin{subfigure}[b]{0.99\columnwidth}  
		\includegraphics[width=\linewidth]{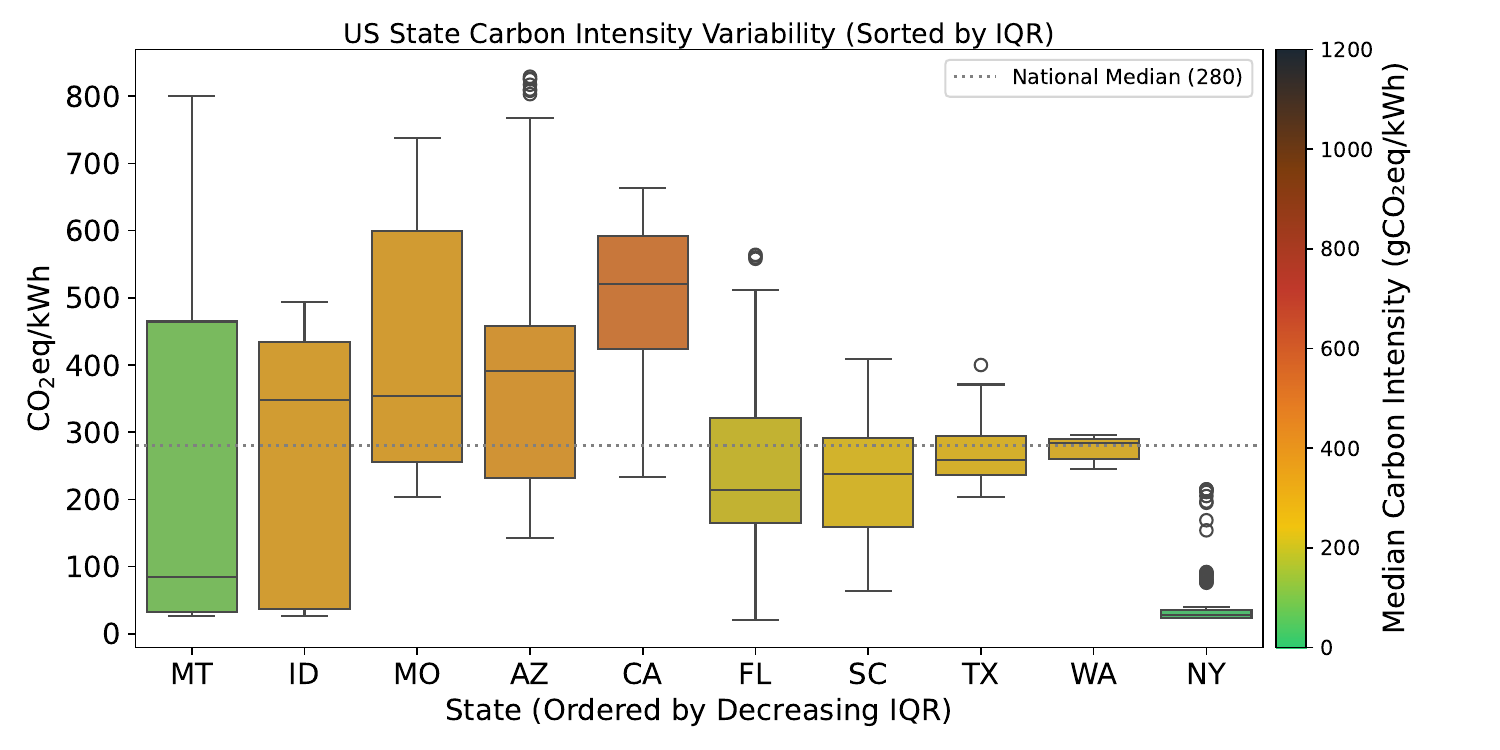}
		\caption{Variability of carbon intensity across regions and over time (24-hour period).} 
	\end{subfigure}
	\caption{Spatial and temporal variations in carbon intensity.}
	\label{fig:carbon_intensity}
\end{figure*}

While there is existing work focusing on the placement of computing tasks and cloud workloads at low-intensity regions and time zones \cite{CADRE, CAFTM, mosaic, CarbonClipper, CarbonContainers, HallCarbonAware}
, most works do not consider the non-negligible carbon emissions during data transfers across datacenters. This paper presents a novel approach to construct the carbon-aware scheduling of data transfer tasks as a Linear Programming (LP) optimization problem and uses standard LP solvers that are efficient and easy to integrate and deploy. More specifically, this paper makes the following contributions:

\begin{itemize}
	\item It introduces a novel carbon-aware data transfer scheduler, LinTS, for inter-datacenter traffic. 
	\item It defines carbon-aware temporal scheduling of data transfers as a linear programming (LP) problem. 
	\item LinTS can make scaling scheduling decisions for transfer requests, unlike common heuristic scheduling algorithms.
	\item LinTS can lower carbon emissions during inter-datacenter transfers by up to 66\% compared to the worst case and up to 15\% compared to other solutions.
\end{itemize}


The paper is organized as follows: Section II provides background into carbon-aware workload shifting strategies and discusses the related work in this area; Section III describes the motivation and foundation for applying LP to carbon-aware transfer scheduling; Section IV presents and discusses evaluations of our LP scheduler in simulation and real-world scenarios; and Section V concludes the paper. 


	\label{intro}
	
	\section{Background and Related Work}

\begin{figure*}[t]
	\centering
	\begin{subfigure}[b]{0.99\columnwidth}  
		\includegraphics[width=\linewidth]{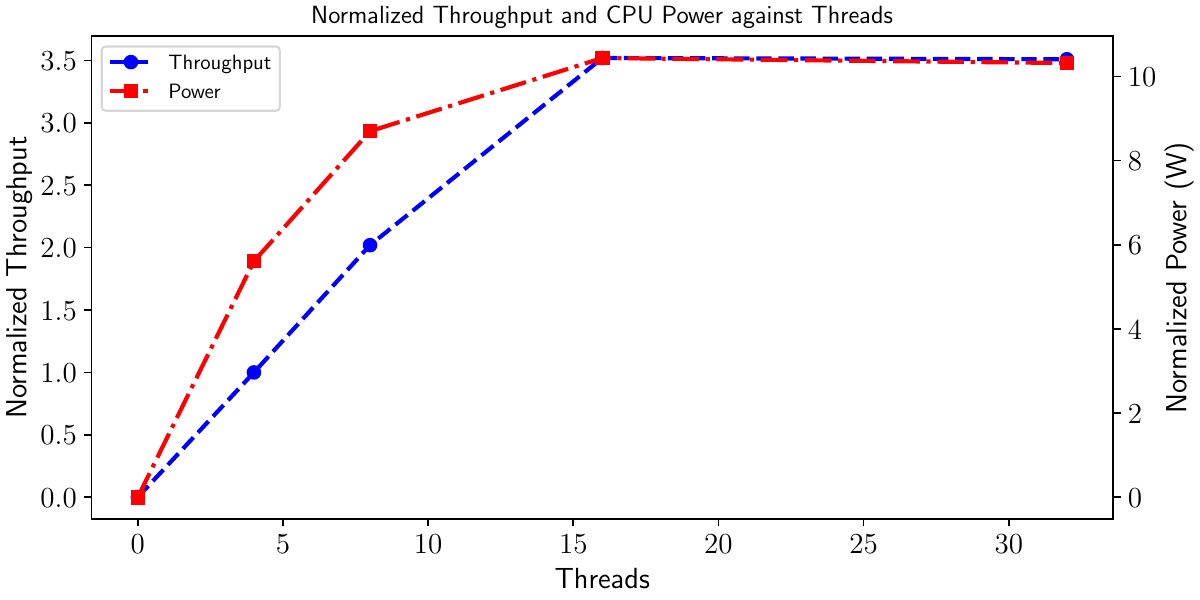}
		\caption{Impact of threads on CPU power demand and throughput.}
	\end{subfigure}
	\hspace{0.04\columnwidth}  
	\begin{subfigure}[b]{0.99\columnwidth}  
		\includegraphics[width=\linewidth]{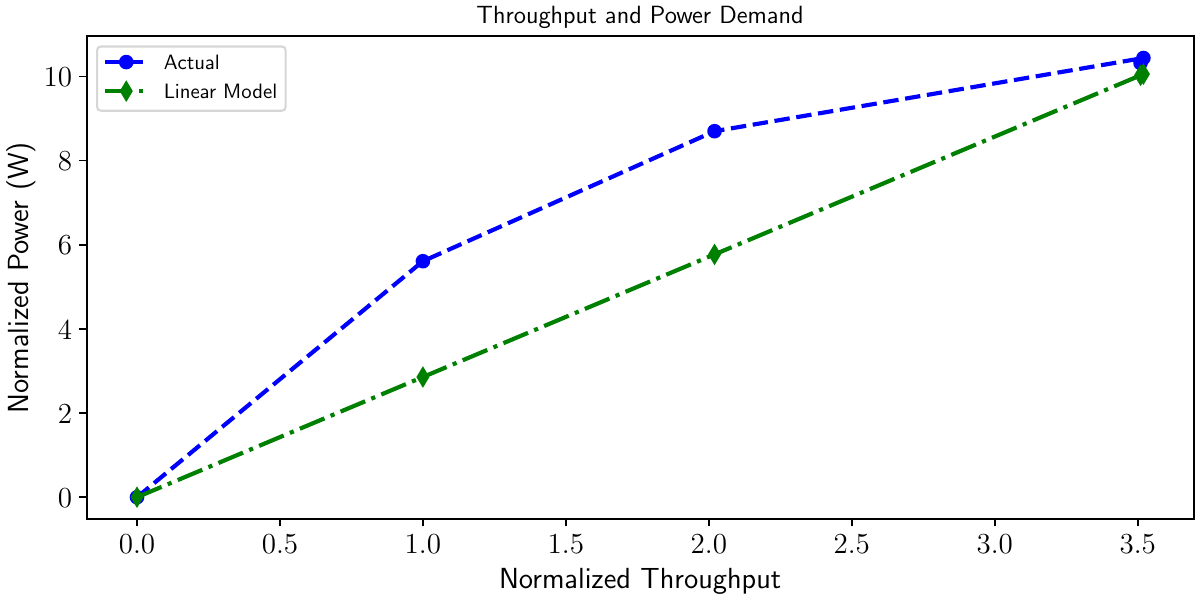}
		\caption{Correlation between the achieved throughput and power demand.}
	\end{subfigure}
	\caption{Modeling the correlation between power demand and achieved throughput.}
	\label{fig:linear_scaling}
\end{figure*}


The spatial and temporal variances in the carbon intensity present an opportunity for minimizing the carbon emission of datacenter workloads by scheduling jobs to geographical locations and at times of the day/week with low carbon intensity through historical data and forecasts, and adjusting task parameters when grid conditions change. As seen in \cite{onTheLimitations}, between 8\% to 31\% in carbon emissions relative to the global average can be made using spatial and temporal approaches and even more when combined. These approaches, however, presume variances in regional grid carbon intensity to exploit and their viability may change as the grid's sources become greener and more uniform. The nature of the workload also impacts how suitable spatial and temporal approaches are. Long-running, data-intensive, batch workloads like machine learning (ML) training are often delay tolerant and interruptible and benefit more from temporal approaches. On the other side, bursty and interactive workloads often benefit more from spatial approaches when user responsiveness and latency are important. 



Carbon intensity for a region can vary over time due to variances in the availability of green sources provided by the grid. For instance, without battery storage, solar power can be highly available on a sunny morning but less so at night, requiring brown sources to substitute for. Similarly, carbon intensity can vary over weeks and months due to seasonal variations. Thus, recurring workloads would need to be carefully managed to minimize carbon emissions year-round. Figure 1(a) shows an example of temporal variations of the carbon intensity of an end-to-end network path, where a workload or network transfer can be scheduled in low-carbon opportunities. 
Some works like \cite{Cucumber} and \cite{CAFTM} tune the power draw of hardware with voltage and frequency scaling, by lowering voltage or frequency in periods of high intensity and increasing them in low-intensity periods. The same scaling principle can be applied to containers, Kubernetes pods, and nodes as seen in \cite{CASA} and \cite{CarbonContainers}. Works like \cite{googleCarbon} and \cite{HallCarbonAware} approach carbon optimization with capacity planning, while $WaitAwhile$ \cite{WaitAwhile} uses a scheduler that defers, interrupts, and resumes workloads when possible. Our work extends the ideas found in $WaitAwhile$ and $CarbonScaler$ \cite{CarbonScaler} for carbon-aware network transfer temporal scheduling.

Among spatial strategies for carbon optimization, most commonly fall into the category of scheduling and real-time migration and often mix these two ideas in their implementations. As a workload arrives, a scheduler can use historical data or carbon intensity forecasts of various datacenters and schedule that workload in the greenest region possible, balancing factors such as deadlines, latency requirements, and the overall load. Figure 1(b) shows the variability of carbon intensity across different regions and over time (24-hour period). In some implementations, a monitoring service watches for large deviations in real-time carbon and load data while this workload is executed and may migrate the task or redirect the client to another datacenter \cite{googleCarbon, CFR}. A variant of this strategy involves sharing a workload among multiple datacenters if it permits, allowing one to better balance carbon efficiency, datacenter load, and availability \cite{CLIMP}. CADRE is a carbon-aware replication planner that makes spatial decisions and places replicas in low-intensity regions \cite{CADRE}. Workloads that are WAN-intensive can benefit from carbon-aware routing and overlay networks, though it is harder to measure this strategy's impact since the routers, switches, and other intermediary nodes along the path must be accounted for. 

All of these strategies require a way to measure or estimate power consumption and a source for carbon intensity data either from power grid providers or through third-party services like $Electricity Maps$ and $WattTime$. A workload may sometimes be more CPU-intensive, I/O-intensive, or network-intensive so some works also seek to isolate the power consumption of a node's component for better accuracy. Power measurement of the CPU package can often be done in software by reading model-specific registers (MSR) or through models with tools like $perf$ \cite{perfWiki}, RAPL \cite{IntelRAPL}, NVML \cite{NVML}, FECoM \cite{FECoM}, and $powerletrics$ \cite{Powerletrics}.

While lowering carbon emissions is the main goal for many of these strategies, it is also important to ensure that datacenter performance, availability, and reliability are not compromised. Thus, performance metrics like throughput, tail latency, scalability, datacenter load, service level agreement (SLA) violations, and cost of operations are often measured to show any effects and tradeoffs of these green strategies \cite{PeelingBack}. Likewise, due to the varying nature of datacenter workloads, some green strategies target workloads with high delay tolerance while others focus on low tolerance workloads.

Most of the carbon-aware works discussed here study datacenter workload scheduling and management and use strategies that are transferable to network data transfer scheduling, but there are a few important distinctions between the problems. Unless the compute task is geographically distributed, most datacenter tasks discussed in these works are executed in one location and only have to account for the carbon intensity of that region, while data transfers will often span multiple hops across different regions, requiring consideration of the carbon intensities of multiple zones at once. Another important distinction is the lack of control over resources in network transfers. Datacenters have control over all the resources needed to execute the task and can tune those resources appropriately, while users often do not enjoy control over routing decisions and intermediate nodes of a data transfer. Additionally, the data transfers can be affected by background data traffic activity, making potential capacity planning and throughput predictions very challenging.



Several works on datacenter workload management can be found that employ spatial strategies, temporal strategies, or both. Some works are concerned primarily with scheduling of tasks and use static optimization techniques with power models, carbon and demand forecasts, or power models to produce execution plans with sometimes mechanisms to reevaluate plans when forecast errors exceed a threshold \cite{mosaic, googleCarbon, CLIMP, WaitAwhile, GAIA}. Other works instead focus on real-time optimization, using admission control strategies \cite{Cucumber}, hardware voltage and frequency scaling \cite{Ecovisor, CAFTM, CASA, Cucumber}, horizontal scaling \cite{CarbonScaler, HallCarbonAware}, reinforcement learning agents \cite{CFR}, and migration \cite{arslan2018high, kosar2005data, kola2004disc, GreenCourier, Caribou}.

Like LinTS, works like \cite{AldossaryMILP}, \cite{CAFTM}, \cite{CarbonClipper}, and \cite{CarbonEdge} use some form of linear, quadratic, and constraint programming in their schedulers with great degrees of success. However, none of these works consider the carbon cost of data transfers between cloud datacenters. While the works discussed above focus on datacenter workloads without consideration of inter-datacenter communication, as seen in replication and distributed storage applications, for example, LinTS explores carbon optimization for workloads with significant network transfers between datacenters using the carbon intensity of the network path itself.

	\label{background} 
	
	\section{LinTS Overview}
	We introduce the data transfer temporal scheduling problem here, where a scheduler must assign time slots and set threads, if possible, for a set of transfer requests with file sizes $\mathcal{J} = \left\{J_1, J_2, \ldots, J_n\right\}$ and deadlines $\mathcal{D} = \left\{D_1, \ldots, D_n\right\}$ while minimizing the total carbon emissions of these transfers. This entails accounting for all nodes along the transfer path. To make carbon-aware decisions, the scheduler can use historical carbon intensity traces or forecasts.

\begin{table}[h]
	\centering
	\scriptsize
	\strutlongstacks{T}
	\tabcolsep=0.21cm
	\begin{tabular}{c l}
       \hline

		\textbf{Notation}& \textbf{Description}\\ 
		\hline
		$\rho_{i, j}$& Throughput of request $i$ at time slot $j$\\
		$\vec{\rho}$& Throughput of all requests at all slots as a flattened vector\\
		$L$& First-hop bandwidth limit of the path\\
		$s_\rho$& Throughput scale constant\\
		$s_P$& Power scale constant\\
		$c_{i,j}$& Combined carbon intensity of request $i$ at time slot $j$\\
		$\mathbf{c}_i$& Combined carbon intensity vector of request $i$\\
		$J_i$& Size of transfer request $i$ in bytes\\
		$D_i$& Deadline of request $i$ in number of slots from origin\\
		$\Delta_\tau$&Length of time slot in seconds\\
		\hline
	\end{tabular}
	\caption{The list of notations used in the paper.}
	\label{tab:notation}
\end{table}

\subsection{LinTS Linear Programming Model}
Linear Programming (LP) is an optimization technique used to minimize or maximize a linear objective function constrained to a set of linear inequalities. This can be expressed in the standard form where one solves for $\mathbf{x}$:
\begin{align*}
	\text{maximize}\quad &\mathbf{c}^T\mathbf{x}\\
	\text{subject to}\quad &A\mathbf{x}\preceq\mathbf{b},\\
	\text{and}\quad &\mathbf{x}\succeq\mathbf{0},
\end{align*}
\noindent where $A$ is a matrix that expresses the linear inequality and $\mathbf{c}$ is the cost vector to optimize for \cite{Chvatal1983Linear}. Thus, to apply linear programming, a linear cost function is needed. Works that study the relationship between network traffic and energy consumption commonly use either a non-linear model, a linear model, or a state-based model where power consumption increases in steps at certain throughput levels \cite{energyIsmail}; for carbon-aware temporal transfer scheduling, we adopt the linear model in this work. For this work, we assume that all requests received by the scheduler are delay-tolerant, interruptible, and schedulable.

Our solution, the \textbf{Lin}early-optimized \textbf{T}ransfer \textbf{S}cheduler (LinTS), relies on the assumption that network throughput through wide-area networks (WANs) and CPU power demand have a linear-like relationship when throughput is less than the bottlenecked capacity. 
To compare the linear model to the observed relationship, transfers are run on Chameleon Cloud~\cite{ChameleonCloud} from TACC in Texas to Chicago. We scale the number of threads (and sockets) exponentially from 4 threads to 32 threads and measure the achieved throughput and corresponding power consumption. As seen in Figures 1(a) and 1(b), while throughput and power draw have a non-linear relationship with the number of threads and with each other, a linear model can be used to approximate the relationship between throughput and power demand while the network path is not saturated or congested. 

If $L$ is the bandwidth limit of the path and $s_{\rho}$ the throughput scale, we model the throughput $\rho$ achieved with $\theta$ threads with the following equation:
\begin{align}
	\rho(\theta) = L\left(1 - \frac{1}{s_{\rho}L\theta + 1}\right)
\end{align}
Similarly, given the maximum power $P_{\text{max}}$, minimum power $P_{\text{min}}$, and power scale $s_P$, we model the CPU power $P$ drawn with $\theta$ threads as follows:
\begin{align}
	\Delta_P &= P_{\text{max}} - P_{\text{min}}\\
	P(\theta) &= \Delta_P \left(1 - \frac{1}{s_P\Delta_P\theta+1}\right) + P_{\text{min}}
\end{align}
While linear programming is useful for choosing times to start, interrupt, and resume transfers, the power and throughput models in Equations 1 and 3 can be used to extend the scheduler to make thread scaling decisions for additional savings in carbon intensity similar to \cite{CarbonScaler}. Linear program solutions cannot be restricted to integers, so the scheduler cannot directly place threads in time slots. Instead, LinTS makes decisions on the required throughput for each slot and then uses the inverse of Equation 1 to convert throughput to threads:
\begin{align}
	\theta(\rho) = \frac{1}{Ls_P}\left(\frac{\rho}{L-\rho}\right)
\end{align}
Substituting Equation 4 into Equation 3, we get the following relation between power draw and throughput:
\begin{align}
	K &= \frac{s_P\Delta_P}{s_\rho L}\\
	P(\rho) &= P_{\text{max}} + \frac{\Delta_P(\rho - L)}{(K-1)\rho + L}
\end{align}
where $K$ is a constant. Restricting throughput to $0\leq\rho\leq L$, we can linearize the equation:
\begin{align}
	P(\rho) = \frac{\Delta_P}{L}\rho + P_{\text{min}}
\end{align}
Expressing Equation 7 in vector terms where a vector element corresponds to a time slot, we get the following linear objective function:
\begin{align}
	\mathbf{c}_{i}^T\mathbf{p} = \frac{\Delta_P}{L}\mathbf{c}_{i}^T\vec{\rho} + \mathbf{p}_{\min}
\end{align}
where $\mathbf{c}_i$ is the carbon intensity vector for job $i$. Equation 8 implies that carbon emissions can be minimized by minimizing throughput, but it does not capture the fact that slower transfers take longer to complete and can often increase overall carbon emissions; this can be accounted for with linear constraints.
\subsection{LinTS Constraints}
The following constraints are required to produce a feasible plan for a given set of transfer requests. Let $\rho_{i, j}$ be the throughput of a request $i$ at time slot $j$, also written as a flattened throughput vector $\vec{\rho}$ in the following constraints.



\noindent\textbf{Deadline constraint.} Each request has a deadline $D$ without slack. Although this cannot be expressed as an inequality that an LP solver can use, we can encode deadline constraints through the dimensions of the throughput vector.
\begin{align*}
	\dim{\vec{\rho}} = \sum_{i} D_i
\end{align*}
\noindent\textbf{Time-slot constraint.} This constraint ensures that the LP solvers allocate enough throughput and time slots to complete the transfer request of size $J$. If a time slot is of length $\Delta_\tau$, then for each request $i$ and time slot $j$,
\vspace{-2pt}
\begin{align*}
	&J_i \leq \sum_{j=1}^{D_i}t_{i,j}\cdot\rho_{i,j} \quad\forall 1\leq i \leq n\\
	&\text{where}\quad t_{i,j} = \begin{cases}
		0 & \text{if slot $j$ not part of request } i\\
		\Delta_\tau & \text{if slot $j$ part of request } i
	\end{cases}
\end{align*}

\noindent\textbf{Thread-limit constraint.} This constraint ensures that the sum of all request bandwidth allocated at a time slot does not exceed the bandwidth limit $L$. Then, for each slot $j$,
\vspace{-2pt}
\begin{align*}
	\sum_{i=1}^{n}\delta_{i,j}\cdot\rho_{i, j} &\leq L\quad\forall 1\leq j \leq \max_i{\left(D_i\right)}\\
	\text{where}\quad\delta_{i,j} &= \begin{cases}
		0 & \text{if request $i$ not part of slot } j\\
		1 & \text{if request $i$ part of slot } j
	\end{cases}
\end{align*}

\noindent\textbf{Input constraint.} Since LinTS assumes a bandwidth limit due to bottlenecks, the throughput constraint $0\leq\rho_{i, j}\leq L$ ensures that the throughput cannot exceed this limit.

Taken together, the temporal scheduling problem can be expressed as the following linear program.
\begin{align*}
	\text{minimize}\quad & \sum_{i=1}^n \sum_{j=1}^{D_i} c_{i,j}\cdot \rho_{i, j}\\
	\text{subject to}\quad &J_i \leq \sum_{j=1}^{D_i}t_{i,j}\cdot\rho_{i,j}\\
	\text{and}\quad &\sum_{i=1}^{n}\delta_{i,j}\cdot\rho_{i, j} \leq L\\
	\text{and}\quad &\dim{\vec{\rho}} = \sum_{i=1}^n D_i\\
	\text{and}\quad &0\leq\rho_{i,j}\leq L \\
	&\forall 1\leq i \leq n, 1\leq j \leq \max_i{\left(D_i\right)}
\end{align*}

\subsection{LinTS Implementation}
LinTS is implemented in Python using SciPy's efficient \texttt{linprog} solver \cite{scipyLinprog}, and is simulated in Python using historical carbon intensity traces. It is designed to integrate with data transfer services as a Python library or a REST API with Flask.

\begin{algorithm}
	\caption{LinTS Algorithm}
	\begin{algorithmic}[1]
		
		\STATE $\text{forecast\_sums} \gets \sum \text{weights}\cdot\text{forecast}$
		\FOR{$f$ in forecast\_sums}
		\STATE $f\gets \text{ExpansionMatrix}\cdot f$
		\ENDFOR
		\STATE $\mathbf{c}\gets \text{flatten}(\text{forecast\_sums})$
		\REQUIRE $\dim{c} = \sum D_i$
		\STATE $A_{ub}\gets \text{[]}$
		\COMMENT{Upper bound linear constraints}
		
		\STATE offset $\gets 0$
		\FOR{$n$ from $0$ to num\_jobs}
		\STATE byte\_sum\_vec $\gets \mathbf{0}$
		\STATE byte\_sum\_vec[i] $\gets \text{slot\_time} \quad\forall \text{offset} \leq i \leq D_n$
		\STATE $A_{ub} \gets \text{append}(A_{ub}, \text{byte\_sum\_vec})$
		\ENDFOR
		
		\FOR{$i$ from $0$ to $\max$(deadlines)}
		\STATE slot\_constraint $\gets \mathbf{0}$
		\FOR{all slots $S$ at time $i$}
		\STATE slot\_constraint[$S$] $\gets 1$
		\ENDFOR
		\STATE $A_{ub}\gets \text{append}(A_{ub}, \text{slot\_constraint})$
		\ENDFOR
		\STATE $b_{ub} \gets -8\cdot \text{data\_size\_vec}$
		\STATE $b_{ub}\gets \text{append}(b_{ub}, \text{target\_thrpt\_vec})$
		\STATE $\text{thrpt\_plan}\gets \text{linprog}(\mathbf{c}, A_{ub}, b_{ub}, (0, L))$
		\STATE $\text{thrpt\_plan}\gets \text{unflatten}(\text{thrpt\_plan})$
		\STATE $\text{thread\_plan}\gets \theta(\text{thrpt\_plan})$
		\RETURN thread\_plan
	\end{algorithmic}
\end{algorithm}

Algorithm 1 is the core workflow of LinTS. Lines 1 to 5 prepare the cost vector from the weighted trace sums and encode the deadline constraint through dimensions. Lines 6 to 12 and 20 construct the deadline constraint, and lines 13 to 19 and 21 prepare the byte constraint. Finally, lines 22 to 24 call the ScipPy LP solver, unwrap the solution, and convert it to threads with Equation 4. Given these constraints, LinTS allows multiple transfer requests to share a time slot as long as their collective throughput does not exceed the bandwidth limit and is free to scale transfer threads up or down as needed.

After producing a plan, the simulator estimates its carbon emissions with the carbon intensity trace and power curve seen in Equation 3. Noise is added to the trace to emulate possible errors in carbon forecasts before iterating over the plan. If a slot has no threads allocated for any request, then the simulator assumes no energy consumption at that time as we want to measure only energy consumed by the transfer requests.
	\label{overview}
	
	
	\section{Evaluation}
	In this section, we evaluate our solution, LinTS, in reducing the carbon emissions of data transfers across regions, comparing it to known scheduling algorithms.

\begin{figure*}[t]
	\centering
	\includegraphics[width=1.1\linewidth]{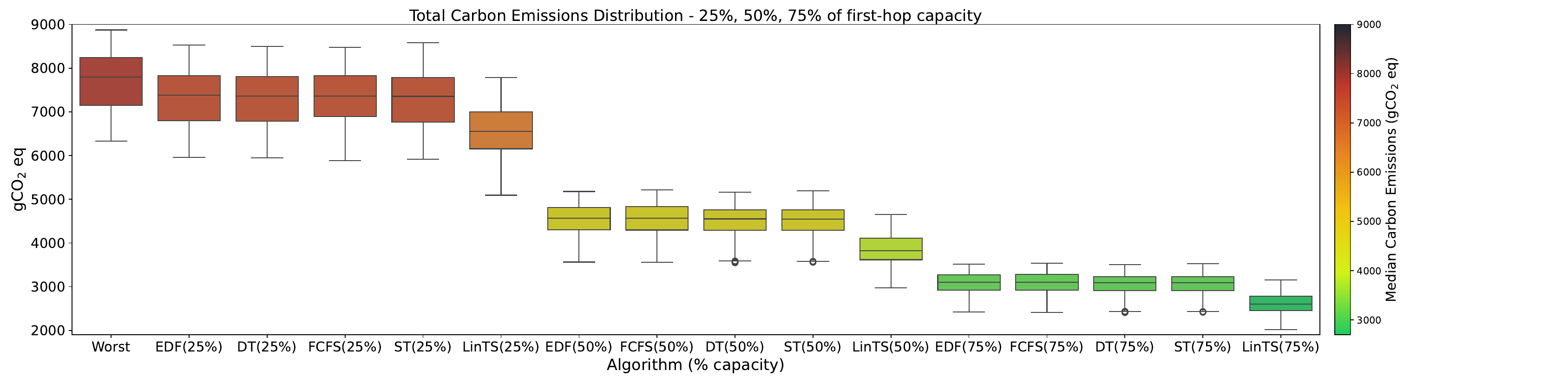}
	\caption{Comparison of algorithms' carbon emissions when restricted to 25\%, 50\%, and 75\% of first-hop capacity.}
	\label{fig:comparison_all}
\end{figure*}

\subsection{Experimental Setup}
\noindent\textbf{Carbon intensity traces.} We use 72-hour slices of historical carbon intensity data from $Electricity Maps$ consisting of hourly measurements for every power zone in the US for all of 2024 \cite{ElectricityMaps}. We pick sites with the highest variability in carbon intensity, namely zones in New Mexico, Colorado, Utah, Wyoming, South Dakota, South Carolina, and Montana as seen in Figure 1(b). While $Electricity Maps$ provides carbon intensity forecasts that would commonly be used in a scheduler, it is currently limited to 0 -- 72 hours depending on the subscription plan,  making it infeasible for large data transfers spanning over several days or with generous deadlines. 
Accounting for errors with forecasts, we add random noise of 5\% and 15\% to our historical traces. LinTS in its current form does not adjust plans as network and carbon conditions change, leaving this capability for future work.

\noindent\textbf{Node and data transfer characteristics.} We simulate a data transfer over up to 8 nodes: a source, up to six intermediate nodes (i.e., router, switch, repeater, etc.), and a destination. While multiple routers and switches are often found in the data transfer path, we choose a simple WAN of up to 8 nodes connected by a long wired network to simplify the simulation. The bandwidth of the link between the source and destination is known and fixed to 1 Gbps, which we call the `first-hop bandwidth', but the capacity of other links is unknown. Since one can expect bottlenecks in the data path over a WAN, the plans produced by the algorithms assume a bottleneck expressed as a percentage of the first-hop bandwidth. Network throughput is assumed to scale non-linearly with threads as seen in Figure 2(a). Similarly, the power consumption of the node is assumed to scale non-linearly with threads and can range from 88W to 100W.

\noindent\textbf{Algorithm configurations.} Our evaluation compares LinTS to heuristic algorithms described in \cite{HanafyDataAlgorithm} and our best heuristic algorithm where they all produce plans for a set of transfer requests with paths and deadlines defined. These plans are evaluated in our simulator to calculate their carbon emissions. All of the heuristic algorithms below assign the highest number of threads allowed by the request's bottleneck to its time slots; this is sufficient since elapsed time is typically the dominant component of a request's footprint. Let $J$ be the number of jobs, $S$ the number of time slots, and $L$ the bandwidth limit.

\begin{figure*}[t]
	\centering
	\includegraphics[width=\linewidth]{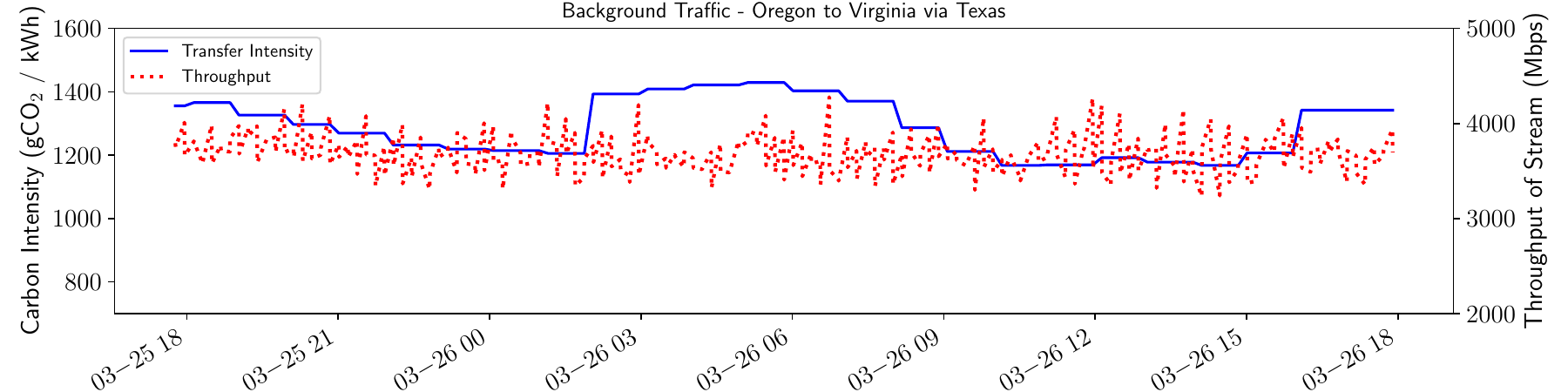}
	\caption{Throughput and carbon intensity of transfers from AWS US-West-2 to US-East-1 via TACC over a 24-hour period.}
	\label{fig:aws-bg}
\end{figure*}

\begin{enumerate}	
	\item\textit{First-come First-serve (FCFS) --} This is the default scheduling algorithm for most file transfer services. As the transfer requests arrive over a network, they are arranged in a queue in the order determined by arrival time. Then, for each request in this order, FCFS simply schedules it without optimizing for carbon footprint by assigning the first $S$ time slots where $S$ is the minimum number of slots needed to complete the request before its deadline; this repeats until the queue is empty. 
	\item\textit{Earliest-Deadline First (EDF) --} The EDF algorithm does not optimize for carbon footprint. The transfer requests are sorted by deadline in ascending order to determine priority. Then, a request with the earliest deadline is assigned to the first $S$ time slots where $S$ is the minimum number of slots needed by the request. The algorithm then picks the request with the next earliest deadline and repeats this process. 
	\item\textit{Worst-Case --} When the transfers are scheduled in a carbon-agnostic manner, in the worst-case scenario, the transfers can potentially be scheduled to the time slots with the highest carbon intensity.  To emulate this case and establish a baseline, we use the EDF algorithm to schedule requests at time slots with the highest carbon intensities. Then, plans are generated randomly and the worst-performing plan between the two methods is used as the worst case.
	\item \textit{Single Threshold (ST) --} This algorithm uses one carbon intensity threshold to allocate time slots to a data transfer. First, the transfer requests are sorted in ascending order of deadline to determine priority. If the carbon intensity at a point in time falls below this threshold, then ST blocks that time slot and allocates it to the request. This continues until the request has sufficient time slots to complete before its deadline. Then, ST moves to the next request and repeats this process. In our implementation, the optimal threshold is found through a binary search since the plan with the lowest feasible threshold is ideal. 
	\item \textit{Double Threshold (DT) --} Instead of one threshold, DT uses a high and low threshold to offset the overhead delay of resuming transfers with lower carbon intensity. Like ST, the requests are sorted by deadline. At a time slot, if the request was been paused at the previous slot and the carbon intensity is below the high threshold, then the slot is given to the request. If the request was paused in the previous slot and the carbon intensity is below the low threshold, then the slot is given to the request. We set $\alpha$, the difference between the thresholds, to 50 and used binary search to find the optimal thresholds. 
	\item \textit{LinTS --} The request size, deadline, and carbon intensity are written as linear constraints and fed to a solver to produce a throughput plan over 72 hours. This plan is then converted to the corresponding threads plan using Equation 4. In our evaluations, we set the bandwidth limit $L$ to 0.25 Gbps for 25\%, 0.5 Gbps for 50\%, and 0.75 Gbps for 75\%. All evaluations use a throughput scale $s_\rho$ of $1/24$. SciPy's default solver switches between the simplex and interior-point methods depending on the size of the input and constraints. 
\end{enumerate}



\noindent\textbf{Transfer requests.} The algorithms above are used to schedule 200 transfer requests received and queued at roughly the same time; this point in time is the origin and we set $t=0$. The file sizes range from 10 GB to 50 GB with deadlines ranging from 48 to 71 hours from origin.

\noindent\textbf{Simulator.} Given a request's path of a source, intermediate node, and destination, the 72-hour carbon intensity traces for their regions are read, and divided and expanded into 288 time slots, 15 minutes each. The request path's intensity is then calculated as the sum of the traces; since we assume all nodes in the path are equally affected by network transfers, we assign equal weight to these nodes. The transfer requirements and traces are then fed into the algorithms above to produce thread plans. The CPU power demand of each plan is then estimated using Equation 3. For our evaluations, $P_{\text{max}} = 100\text{W}$ and $P_{\text{min}} = 88\text{W}$. The power scale $s_P$ is set to $1/50$. Energy consumption and carbon emissions are calculated using the traces. Noise is added to the traces to model errors in forecasts.


\noindent\textbf{Evaluation metric.} Since all of the evaluated algorithms produce feasible plans given transfer sizes and deadlines, we compare the total emissions and resiliency of the algorithms.

\begin{table}[t]
	\centering
	\scriptsize
	\strutlongstacks{T}
	\tabcolsep=0.21cm
	\begin{tabular}{|c|c|c|c|c|c|c|}
		\hline
		\Longunderstack[c]{Bandwidth\\Limit}& \Longunderstack[c]{Worst\\Case}& \Longunderstack[c]{Earliest\\Deadline}& \Longunderstack[c]{FCFS}& DT&ST & \textbf{LinTS}\\ 
		\hline
		25\%&\multirow{3}{*}{7.14 kg} &6.75 kg &6.76 kg & 6.75 kg &6.74 kg &\textbf{6.08 kg} \\
		\cline{0-0}\cline{3-7}
		50\%&&4.12 kg &4.11 kg &4.09 kg&4.09 kg&\textbf{3.56 kg} \\
		\cline{0-0}\cline{3-7}
		75\%&&2.80 kg &2.79 kg &2.77 kg&2.77 kg&\textbf{2.42 kg} \\
		\hline 
	\end{tabular}
	\caption{Average carbon emissions of algorithms at 25\%, 50\%, and 40\% of the first-hop bandwidth with 5\% error. The best-performing algorithm is shown in bold.}
	\label{tab:5p_error}
\end{table}

\subsection{Evaluating LinTS}
The potential and differences of these algorithms are highlighted best when there is high variability in carbon intensity over time. The worst-case emission is computed by comparing emissions of the worst heuristic and random solution searches and taking the larger of the two values. Table II shows the total carbon emissions of the algorithms when restricted to 25\%, 50\%, and 75\% of the first-hop bandwidth (1 Gbps) with 5\% noise added to the traces when evaluated, and Table III shows total carbon emissions for each algorithm but with 15\% noise added instead; in both scenarios, LinTS outperforms the other algorithms with 10.1\% lower carbon emissions at 25\% capacity, 14.2\% lower emissions at 50\% capacity, and 15.4\% lower emissions at 75\% capacity when compared to FCFS, averaging results from the 5\% and 15\% error scenarios. Compared to the worst-case, LinTS achieves 14.8\%,  50.1\%, and 66.1\% lower emissions at 25\%, 50\%, and 75\% capacity respectively on average.

LinTS outperforms both ST and DT in all capacity settings with 9.8\%,  13.6\%, and 13.5\% lower emissions respectively on average. Figure 3 highlights these differences with the distribution of emissions of each algorithm with 15\% noise added and capacity limited to 25\%, 50\%, and 75\% of first-hop bandwidth respectively. The solutions generated by LinTS in general save more carbon than the heuristic algorithms as indicated by the lower median and quartiles, given that LinTS allows multiple jobs to run in a time slot and that it makes scaling decisions with threads unlike the other algorithms here.

\begin{table}[t]
	\centering
	\scriptsize
	\strutlongstacks{T}
	\tabcolsep=0.21cm
	\begin{tabular}{|c|c|c|c|c|c|c|}
		\hline
		\Longunderstack[c]{Bandwidth\\Limit}& \Longunderstack[c]{Worst\\Case}& \Longunderstack[c]{Earliest\\Deadline}& \Longunderstack[c]{FCFS}& DT& ST & \textbf{LinTS}\\ 
		\hline
		25\%&\multirow{3}{*}{7.69 kg} &7.30 kg &7.30 kg &7.29 kg &7.28 kg & \textbf{6.56 kg} \\
		\cline{0-0}\cline{3-7}
		50\%&&4.51 kg &4.52 kg &4.48 kg &4.48 kg &\textbf{3.84 kg} \\
		\cline{0-0}\cline{3-7}
		75\%&&3.06 kg &3.07 kg &3.04 kg &3.04 kg &\textbf{2.61 kg} \\
		\hline 
	\end{tabular}
	\caption{Average carbon emissions of algorithms at 25\%, 50\% and 75\% of the first-hop bandwidth with 15\% error. The best-performing algorithm is shown in bold.}
	\label{tab:15p_error}
\end{table}

\subsection{Impact of Scheduling on Performance}
In a WAN setting, it is possible for two or more transfers to share links along their paths, which can lead to link contention and congestion. Thus, background network activity can negatively affect the capacity and bottleneck of the transfer path at any point in time. With enough demand along the path at time $t$, a scheduler that starts or resumes a transfer at $t$ can potentially increase its total carbon emission since it may slow down and need more time to complete; deadline SLAs may be violated too. If congestion improves when the transfer is scheduled to start or resume, it may finish faster and yield higher carbon savings. Our evaluations above do not account for these possibilities since all of the tested scheduling algorithms do not measure or predict network congestion. While setting a conservative bottleneck capacity can help construct plans resilient to congestion, this is not necessarily a foolproof solution for highly stochastic networks. 

Figure \ref{fig:aws-bg} shows the carbon intensity and throughput of $Iperf$ transfers \cite{iperfTCP} from AWS US-West-2 in Oregon to AWS US-East-1 in Virginia through TACC in Texas serving as the intermediate node over a 24-hour period. $Iperf$ is run repeatedly with a fixed number of sockets to assess any significant background traffic and potential congestion between these sites, and throughput is measured every 10 minutes. Hourly carbon intensity data is collected at each hop's location, starting from Oregon, then Washington, Texas, Georgia, New York, New Jersey, and finally Virginia; these locations are determined through the traceroute utility.  This intensity data is then combined into a single intensity trace of the path as a weighted sum.

In our case, the throughput varies from 3.2 to 4 Gbps over 24 hours, which can adversely affect plans produced by any scheduler. As AWS allows users to temporarily boost throughput for a price, it is possible for a surge in network activity to cause congestion and further hurt performance. Considering these changes in network characteristics and their impact on transfer scheduling, a forecast or monitoring service is needed to predict or measure background activity and reevaluate the schedule which is often seen in some datacenter workload scheduling-related work \cite{HallCarbonAware}. This is a potential research extension left for future work.
	
	\section{Conclusion and Future Work}
	We present LinTS, a scheduler that intelligently schedules and scales data transfers between cloud datacenters by constructing a linear optimization problem with carbon intensity forecasts while respecting request deadlines and requirements. Our simulations show LinTS outperforming threshold methods commonly seen in other works while remaining lightweight and easy to integrate with transfer services. With additional constraints, LinTS can be extended for spatiotemporal scheduling, and multi-objective solvers can be used when there are tradeoffs in the objective function.
	\label{conclusion}
	
	\section*{Acknowledgements}
	This project is in part sponsored by the National Science Foundation (NSF) under award number OAC-2313061. We also thank Chameleon Cloud for making their resources available for the experiments of this work.
	
	\bibliographystyle{IEEEtran}
	\bibliography{references}

	
\end{document}